\documentclass[aps,prl,twocolumn,showpacs,preprintnumbers]{revtex4}

\begin{document}

\preprint{UFIFT-QG-08-04}

\title{Cosmology Is Not A Renormalization Group Flow}

\author{R. P. Woodard}
\email{woodard@phys.ufl.edu}
\affiliation{Department of Physics, University of Florida,
             Gainesville, FL 32611, USA}

\begin{abstract}

A critical examination is made of two simple implementations of the
idea that cosmology can be viewed as a renormalization group flow.
Both implementations are shown to fail when applied to a massless, 
minimally coupled scalar with a quartic self-interaction on a 
locally de Sitter background. Cosmological evolution in this model 
is not driven by any RG screening of couplings but rather by
inflationary particle production gradually filling an initially
empty universe with a sea of long wavelength scalars.

\end{abstract}


\maketitle

\section{Introduction}

The Renormalization Group (RG) describes how the predictions of
flat space quantum field theory and statistical mechanics change when 
all spacetime coordinates are adiabatically scaled up by a constant 
\cite{ZJ},
\begin{equation}
{\rm Renormalization\ Group:} \qquad x^{\mu} \longrightarrow A \times
x^{\mu} \; . \label{RGscale}
\end{equation}
Cosmological evolution can be described by a superficially similar 
rescaling of infinitesimal conformal coordinate intervals \cite{Linde}
\begin{equation}
{\rm Cosmological\ Evolution:} \qquad dx^{\mu} \longrightarrow a(\eta) 
\times dx^{\mu} \; . \label{cosmoscale}
\end{equation}
It is not clear that the one rescaling relates to the other. However,
particle physicists know so much about RG flows, and so little about how 
quantum field theoretic states evolve in an expanding universe, that many 
have posited there is a relation and attempted to exploit it \cite{Many}.

Two simple techniques have been suggested for implementing the RG idea:
\begin{itemize}
\item{{\it Naive Scaling} --- setting the parameter $A$ in (\ref{RGscale}) 
to the ratio of the cosmological scale factor $a(\eta)$ to its initial 
value; and}
\item{{\it Hubble Scaling} --- setting $1/A$ to the ratio of the 
instantaneous Hubble parameter, $H(\eta) \equiv a'/a^2$, to its initial 
value.}
\end{itemize}
Both techniques are dubious. For naive scaling the flat space loop 
amplitudes that determine the various $\beta$ and $\gamma$ functions 
involve integrations over all times. This makes no difference for
constant scalings of the metric --- and introduces no nonlocality, even 
in curved space \cite{wald} --- but it can matter a great deal when the
scaling is made time dependent. {\it Which time ought we to pick, and why?} 
Hubble scaling corresponds to taking $H(\eta)$ as the dimensional 
regularization mass scale $\mu$. That is possible in de Sitter background, 
for which $H$ is a constant, but it would break general coordinate 
invariance otherwise.

The burden of this paper is that both of these scalings are provably 
wrong in a simple theory where their predictions can be checked. No 
shame should attach to having explored the RG idea. Using the RG to 
understand cosmology would have effected a vast simplification. And 
there is still the chance that some more complicated scaling works, 
or even that one of the simple scalings works for other theories. It 
should also be noted that curved space in no way precludes using the
RG conventionally to relate quantities at different constant scales
\cite{Some}. Nor do these comments apply to using the RG on the world
sheet of a string model \cite{Mavro}.

The model for which RG predictions can be checked is a massless, minimally
coupled scalar with a quartic self-interaction on a nondynamical, locally
de Sitter background. Perturbative results are described in Section 2.
These suffice to falsify Hubble Scaling. Section 3 presents the leading 
logarithm solution for the late time limit of this model which was obtained 
by Starobinski\u{\i} and Yokoyama \cite{SY}. This falsifies Naive Scaling. 
Section 4 describes the true origin of evolution in this model.

\section{The Model}

The background geometry is the $D$-dimensional, conformal coordinate patch 
of de Sitter space,
\begin{equation}
ds^2 = a^2 \Bigl[-d\eta^2 + d\vec{x} \cdot d\vec{x}\Bigr] \quad ,
\quad a(\eta) \equiv -\frac1{H \eta} \; \label{background}
\end{equation}
Here and throughout the Hubble constant $H$ relates to the bare 
cosmological constant $\Lambda$ as $H^2 \equiv \Lambda/(D-1)$. 
In terms of the renormalized fields and couplings the Lagrangian is,
\begin{eqnarray}
\lefteqn{\mathcal{L} = -\frac12 (1 \!+\! \delta Z) \partial_{\mu} \varphi 
\partial_{\nu} \varphi g^{\mu\nu} \sqrt{-g} - \frac1{4 !} (\lambda \!+\!
\delta \lambda) \varphi^4 \sqrt{-g} } \nonumber \\
& & \hspace{2cm} - \frac12 \delta \xi \varphi^2 R \sqrt{-g} - 
\frac{(D-2) \delta \Lambda}{16 \pi G} \, \sqrt{-g} \; . \qquad \label{model}
\end{eqnarray}
There is no mass counterterm because the bare mass is zero and mass is 
multiplicatively renormalized. However, even this minimally coupled scalar 
requires a conformal counterterm $\delta \xi$, and there is of course
field strength and coupling constant renormalization. That suffices for 
noncoincident one-particle-irreducible (1PI) functions. Additionally
renormalizing the stress tensor requires the $\delta \Lambda$
counterterm.

The expectation value of the stress tensor has been computed in this
model at one and two loop orders \cite{OW}. The scalar self-mass-squared 
has also been computed at one and two loop orders \cite{BOW}, and used
to quantum-correct the scalar mode functions \cite{KO}. In each case the 
calculations were performed using a version of the Schwinger-Keldysh formalism 
\cite{JSKTMBMLVK}, in the presence of an initial state \cite{FW} which is 
free Bunch-Davies vacuum (for modes which are initially sub-horizon
\cite{AF}) at $\eta = -1/H$.

An unfortunate renormalization scheme was employed involving 
a mass counterterm, which on de Sitter background (\ref{background})
cannot be distinguished from the conformal counterterm for 1PI functions.
However, it is simple to convert to the convention given in (\ref{model}). 
When this is done, the various counterterms can all be expressed in terms 
of the dimensionless coupling constant,
\begin{equation}
\lambdabar \equiv \frac{\lambda}{16 \pi^2} \, \Bigl(\frac{H}{\sqrt{4 \pi}}
\Bigr)^{D-4} \; . \label{lambdabar}
\end{equation}
The lowest nontrivial results for $\delta Z$, $\delta \lambda$ and 
$\delta \xi$ are,
\begin{eqnarray}
\delta Z & = & -\frac{\lambdabar^2}{12} \, \frac{\Gamma^2(1 - \frac12
\epsilon)}{(1 - \frac32 \epsilon) (1 - \epsilon) (1 - \frac34 \epsilon)
\epsilon} + O(\lambdabar^3) \; , \label{dZ} \\
\delta \lambda & = & \lambda \times \Biggl\{ 3 \lambdabar \, \frac{\Gamma(1 -
\frac12 \epsilon)}{(1 - \epsilon) \epsilon} + O(\lambdabar^2) \Biggr\} 
\; , \label{dL} \\
\delta \xi & = & - \frac{\lambdabar}{12} \, \frac{\pi \cot(\frac12 \pi
\epsilon) (1 - \epsilon) \Gamma(1 - \epsilon)}{(1 - \frac13 \epsilon)
(1 - \frac14 \epsilon) \Gamma(1 - \frac12 \epsilon)} + O(\lambdabar^2)
\; . \qquad \label{dX}
\end{eqnarray}
Here and henceforth we define $\epsilon \equiv 4 - D$. Of course
$\delta Z$ and $\delta \lambda$ agree up to finite terms with the well
known results of flat space \cite{ZJ}. The one and two loop results for
$\delta \Lambda$ are,
\begin{eqnarray}
\lefteqn{\frac{(D-2) \delta \Lambda}{16 \pi G} = \frac{H^4}{\lambda}
\Biggl\{ \frac{3 \lambdabar}{2} \, \frac{(1\!-\!\epsilon) (1 \!-\!\frac12 
\epsilon) (1 \!-\!\frac13 \epsilon) \Gamma(1 \!-\! \epsilon)}{(1 - \frac14 
\epsilon) \Gamma(1 - \frac12 \epsilon)} } \nonumber \\
& & - \frac{\lambdabar^2}{2 \epsilon (1 \!-\! \frac14 \epsilon)} \Bigl[ 
\frac{\pi \cot(\frac12 \pi \epsilon) \epsilon (1\!-\!\epsilon) \Gamma(1\!-\! 
\epsilon)}{2 \Gamma(1 - \frac12 \epsilon)} \Bigr]^2 \!\!+ O(\lambdabar^3) 
\Biggr\} .  \quad
\end{eqnarray}

The expectation value of the stress tensor takes the perfect fluid form,
$\langle \Omega \vert T_{\mu\nu} \vert \Omega \rangle = p \times g_{\mu\nu}
+ (\rho + p) \times a^2 \delta^0_{~\mu} \delta^0_{~\nu}$.
It is simplest to express the renormalized pressure and energy density
in units of $H^4/\lambda$,
\begin{eqnarray}
\lefteqn{\frac{\lambda}{H^4} \times p = \lambdabar^2 \Biggl[ -2 \ln^2(a) - 
\frac72 \ln(a) + \frac53 - \frac{\pi^2}3 } \nonumber \\
& & \hspace{2.5cm} - \frac23 \sum_{n=2}^{\infty} \frac{(n\!-\!3)(n\!+\!1)}{
n^2 \, a^n} \Biggr] + O(\lambdabar^3) \; , \qquad \label{justp} \\
\lefteqn{\frac{\lambda}{H^4} \!\times\! (\rho \!+\! p) = \lambdabar^2 \Biggl[ 
-\frac43 \ln(a) - \frac{13}{18} + \frac89 a^{-3} } \nonumber \\
& & \hspace{3.5cm} - \frac23 \sum_{n=2}^{\infty} \frac{(n\!+\!1)}{n \, a^n}
\Biggr] + O(\lambdabar^3) \; . \qquad \label{rho+p}
\end{eqnarray}
It is straightforward to verify that these results obey stress-energy
conservation and that they violate the weak energy condition \cite{OW},
\begin{equation}
\frac{d}{d \eta} \Bigl[a^3 (\rho + p)\Bigr] = a^3 \frac{d p}{d \eta} 
\qquad , \qquad \rho + p < 0 \; .
\end{equation}

Each of the inverse powers of $a$ in (\ref{justp}-\ref{rho+p}) is 
separately conserved. This and their rapid falloff away from the initial
value surface have prompted the speculation that these terms can be absorbed 
into a perturbative correction of the initial state \cite{OW}. Of course 
the constant part of (\ref{justp}) can be absorbed into a finite shift of 
the bare cosmological constant. This leaves the secular contributions,
\begin{eqnarray}
\frac{\lambda}{H^4} \times p_{\rm sec} \!\!\!& = & \!\lambdabar^2 \Bigl[ -2 
\ln^2(a) - \frac72 \ln(a) \Bigr] + O(\lambdabar^3) \; , \qquad \label{psec} \\
\frac{\lambda}{H^4} \!\times\! (\rho \!+\! p)_{\rm sec} \!\!\!& = & \!
\lambdabar^2 \Bigl[ -\frac43 \ln(a) - \frac{13}{18} \Bigr] + O(\lambdabar^3) 
\; . \qquad \label{rhosec}
\end{eqnarray}

We conclude this section by giving the RG predictions for this model.
The RG evolution of its flat space analogue is the best understood of all 
quantum field theories. In fact, it is the paradigm for critical phenomena 
in systems for which the order parameter is one dimensional \cite{ZJ}. And 
there is absolutely no doubt that scaling $A$ to infinity carries this 
system to a free theory in $D=4$. Of course the cosmological scale factor
$a(\eta)$ grows without bound, so the prediction of Naive Scaling is that
all correlation functions should become Gaussian at late times. On the
other hand, the Hubble parameter is constant in de Sitter background, so
the prediction of Hubble Scaling is that the expectation values of 
operators such as the stress tensor should show no change with time. 

{\it The fact that two simple implementations of the RG idea give
different evolutions is already suspicious}, even without explicit results.
Of course the time dependence evident even in perturbative results such 
as (\ref{psec}-\ref{rhosec}) suffices to falsify the prediction of Hubble 
Scaling. The next section will show that the prediction of Naive Scaling 
is also wrong.

\section{Leading Log Solution}

The factors of $\ln(a)$ in expressions (\ref{psec}-\ref{rhosec}) are known 
as {\it infrared logarithms}. Any quantum field theory which involves
undifferentiated gravitons or massless, minimally coupled scalars will show 
similar infrared logarithms in some of its Green's functions. They occur
as well in scalar quantum electrodynamics \cite{PTWP}, in Yukawa theory
\cite{PW2GPMW1}, in pure quantum gravity \cite{TW}, and in quantum
gravity with fermions \cite{MW}. They even contaminate loop corrections to 
the power spectrum of cosmological perturbations \cite{SW1,BSVMSKCBPvdmsUM} 
and other fixed-momentum correlators \cite{SW2}.

Infrared logarithms introduce a fascinating secular element into the usual,
static results of quantum field theory. It was this secular evolution in
perturbative results such as (\ref{psec}-\ref{rhosec}) that allowed us to 
falsify the RG prediction for Hubble Scaling. Achieving a similar 
falsification for Naive Scaling requires a more powerful analysis because 
the continued growth of $\ln(a)$ must eventually overwhelm the loop 
counting parameter $\lambdabar$, which causes perturbation theory to
break down. To evolve to late times requires a nonperturbative technique.

For certain models there are natural resummation schemes such as the $1/N$
expansion \cite{CMBCVHSSRS}. A more general technique is suggested by
the form of the expansion for the pressure. The $-2 \ln^2(a)$ in 
expression (\ref{psec}) is a {\it leading logarithm}, while the $-\frac72
\ln(a)$ is a {\it sub-leading logarithm}. For this model one can show that 
each extra factor of $\lambdabar$ in the perturbative expansion of any 
quantity brings at most two more powers of $\ln(a)$ \cite{PTsW3}. The 
general expansion for the pressure is therefore $H^4$ times,
\begin{equation}
\sum_{n=1}^{\infty} \lambdabar^n 
\Bigl\{\!c_{n,0} [\ln(a)]^{2n} \!+\! c_{n,1} [\ln(a)]^{2n-1} \!+\!
\dots \!+\! c_{n,2n-1} \ln(a)\!\Bigr\} . \label{genform}
\end{equation}
Here the constants $c_{n,k}$ are pure numbers. Perturbation theory 
breaks down when $\ln(a) \sim 1/\sqrt{\lambdabar}$, at which time the 
leading infrared logarithms at each loop order contribute numbers of 
order one. In contrast, the subleading logarithms are all suppressed 
by at least one factor of the small parameter $\sqrt{\lambdabar}$. So a 
sensible approximation is to retain only the leading infrared logarithms.

Starobinski\u{\i} has developed a simple stochastic formalism \cite{AAS}
that reproduces the leading infrared logarithms at each order \cite{RPW3}
for any scalar potential model, including (\ref{model}). Probabilistic 
representations of inflation have been studied to understand initial 
conditions \cite{AVNS}, global structure \cite{GLMLM} and non-Gaussianity 
\cite{RST}. However, the focus here is on recovering the most important 
secular effects of inflationary quantum field theory \cite{SJRSNNWVMMENPR}. 
It is of particular importance that Starobinski\u{\i} and Yokoyama have 
derived the late time limit whenever the potential is bounded below 
\cite{SY}. This is the true analogue of what the RG accomplishes.

Starobinski\u{\i}'s formalism facilitates developing lead\-ing
log\-arithm expansions to very high orders \cite{RPW3},
\begin{eqnarray}
\lefteqn{\langle \Omega \vert \varphi^{2n}(x) \vert \Omega 
\rangle_{\rm leading} = } \nonumber \\
& & \hspace{-.2cm} (2n \!-\!1)!!  \Bigl[\frac{H^2 \ln(a)}{4 \pi^2} \Bigr]^n 
\Biggl\{1 - \frac{2 n^2 \!+\! 2n}{9} \, \lambdabar \ln^2(a) + \nonumber \\
& & \hspace{.5cm} \frac{70 n^4 \!+\! 340 n^3 \!+\! 450 n^2 \!+\! 148 n}{3645}
\, \lambdabar^2 \ln^4(a) - \dots \Biggr\} . \qquad
\end{eqnarray}
It also allows one to compute the late time limits \cite{SY,RPW3},
\begin{equation}
\lim_{a \rightarrow \infty} \langle \Omega \vert \varphi^{2n}(x) \vert 
\Omega \rangle_{\rm leading} = \frac{\Gamma(\frac{n}2 + \frac14)}{
\Gamma(\frac14)} \Bigl[ \frac{9 H^4}{16 \pi^4 \lambdabar}\Bigr]^{\frac{n}2} 
\; . \label{nonpert}
\end{equation}
Now recall that the RG prediction for Naive Scaling is that the late time
limit becomes Gaussian. Were that correct, the expectation value of $2n$ 
coincident fields would be $(2n\!-\!1)!!$ times the $n$-th power of the 
coincident 2-point function. From (\ref{nonpert}) one can see that the
actual result for $n=2$ is smaller by a factor of about .72948. For
$n=3$ the ratio is approximately .437688; for $n=4$ it is about .22806; 
and the ratio falls exponentially for large $n$. Hence Starobinski\u{\i}'s 
formalism falsifies the RG prediction for Naive Scaling.

\section{Conclusions}

We have seen that the evolution of this model is not described by any
simple RG flow. It is driven instead by the inflationary expansion ripping 
long wavelength, virtual scalars out of the vacuum. That process can be 
derived in very simple terms using the energy-time uncertainty principle 
and the scalar's breaking of classical conformal invariance \cite{RPW}. It 
results in a slow growth of the scalar field strength that appears in the
classic result for the coincidence limit of the free scalar propagator
\cite{VFLS},
\begin{equation}
\langle \Omega \vert \varphi^2(x) \vert \Omega \rangle_{\rm free} = 
{\rm Divergent\ Constant} \; + \frac{H^2 \ln(a)}{4 \pi^2} \; .
\end{equation}
This tends to drive the scalar up its $\varphi^4$ potential, which induces 
the negative pressure $\sim \ln^2(a)$ that is evident at lowest order in 
expression (\ref{psec}). At next order the classical force associated 
with being away from the potential minimum tends to decrease the scalar
field strength, which makes the pressure less negative, and so on. The
resulting pressure is an oscillating series of leading logarithms that 
approaches a constant at late times,
\begin{eqnarray}
\lefteqn{-\frac{H^4}{8 \pi^2} \Bigl[\lambdabar \ln^2(a) - \frac43 \, 
\lambdabar^2 \ln^4(a) 
+ \frac{5936}{3645} \, \lambdabar^3 \ln^6(a) - \dots 
\Bigr] } \nonumber \\
& & \hspace{5.5cm} \longrightarrow -\frac{3 H^4}{32 \pi^2} \; . \qquad
\end{eqnarray}

This is just what common sense suggests must happen, although 
Starobinski\u{\i} and Yokoyama deserve enormous credit for having 
proved it \cite{SY}. Eventual equilibrium is inevitable because there 
is no increase in the upward pressure on $\varphi^2$ from inflationary 
particle production, whereas the downward pressure from the classical 
force grows without bound. So evolution in this model can be understood 
in even simpler terms than RG flows. And it was, after all, the hope 
for such enlightenment that motivated the RG approach to cosmology 
\cite{Many}.

\section*{Acknowledgements}

I have enjoyed discussions with S. P. Miao, E. Mottola, T. Prokopec, 
A. A. Starobinski\u{\i} and N. C. Tsamis. This work was partially supported 
by NSF grant PHY-0653085, by the Institute for Fundamental Theory at the 
U. of Florida and by FAPESP and CNPq in Brazil.

\end{document}